[Original Article]

# Probing 10 μK stability and residual drifts in the cross-polarized dual-mode stabilization of single-crystal ultrahigh-$Q$ optical resonators

Jinkang Lim[1,*], Wei Liang[2], Anatoliy A. Savchenkov[2], Andrey B. Matsko[2,*], Lute Maleki[2], and Chee Wei Wong[1,*]

[1] Fang Lu Mesoscopic Optics and Quantum Electronics Laboratory, University of California, Los Angeles, CA 90095 USA

[2] OEwaves Inc., 465 North Halstead Street, Suite 140, Pasadena, CA 91107 USA

* Author email addresses: jklim001@ucla.edu; andrey.matsko@oewaves.com; cheewei.wong@ucla.edu

**Abstract**

The thermal stability of monolithic optical microresonators is essential for many mesoscopic photonic applications such as ultrastable laser oscillators, photonic microwave clocks, and precision navigation and sensing. Their fundamental performance is largely bounded by thermal instability. Sensitive thermal monitoring can be achieved by utilizing cross-polarized dual-mode beat frequency metrology, determined by the polarization-dependent thermorefractivity of a single-crystal microresonator, wherein the heterodyne radio-frequency beat pins down the optical mode volume temperature for precision stabilization. Here, we investigate the correlation between the dual-mode beat frequency and the resonator temperature with time and the associated spectral noise of the dual-mode beat frequency in a single-crystal ultrahigh-$Q$ MgF$_2$ resonator to illustrate that dual-mode frequency metrology can potentially be utilized for resonator temperature stabilization reaching the fundamental thermal noise limit in a realistic system. We show a resonator long-term temperature stability of 8.53 μK after stabilization and unveil various sources that hinder the stability from reaching sub-μK in the current system, an important step towards compact precision navigation, sensing and frequency reference architectures.

**Introduction**

High-precision clocks and oscillators are cornerstones for global navigation to provide accurate timing and distance information, via satellite-to-satellite or satellite-to-ground communications, for the synchronization of electronic device communications and for quantum information processing using atomic qubits [1-3]. Likewise, precision time-frequency transfer in timing links is helpful for synchronization and data exchange in coherent arrays, gravitation sensing and relativity measurements [4]. Current remote free-space links relying on radio-frequency communications have limited data transfer rates largely due to the available spectral bands, and often require large antenna sizes and high power consumptions due to diffraction, which naturally demands higher frequency carriers. However, the oscillators based on electronics attain their spectral purity from the high-quality factor of the resonator circuit, which usually degrades with increasing frequency. This design, therefore, potentially increases not only the size and the power consumption but also the complexity of the signal generation subsystems. As a result, free-space low-noise optical carrier networks provide an alternative with compact footprints for high-precision timing synchronization and high data rate links [4-6].

Optical microresonator devices are a potential platform to provide both high spectral purity and stable clock operation for future timing networks. High quality factor ($Q$) solid-state microresonators, especially whispering-gallery resonators [7], have demonstrated superior performances in ultralow-noise laser oscillators and microwave generation via optical frequency division [8-13]. They can also be disciplined to optical atomic transitions in analogous architectures of quartz-crystal microwave oscillators disciplined to radio-frequency atomic hyperfine clock transitions [14]. Single-crystal whispering-gallery resonators are advantageous when compared to other solid-state microresonators since they have $Q$s more than a billion, significantly higher than those of quartz oscillators ($10^6$) and rubidium clock hyperfine atomic transitions ($10^7$) used in current time and frequency standards. This high-$Q$ or narrow-linewidth feature provides lower phase noise and thus potentially lowers the timing jitter of the electromagnetic carrier. Furthermore, the small form factor of the resonators also allows for miniaturization towards space applications [15]. However, solid-state microresonators experience large thermal noise induced by the medium, potentially hindering the ultimate performance of the microresonators towards advanced applications in ultrastable laser oscillators and photonic microwave clocks. Temperature-induced aging could also be detrimental for sensitive optical sensors using the microresonators. For instance, the achievable bias stability of the microresonator resonant gyroscopic sensor is typically larger than 3° hr$^{-1}$ [16] and is bounded by the resonator thermal stability.

For sub-millimeter-scale resonators, theoretical models predict that sub-μK stability is required for achieving the thermodynamically limited noise level corresponding to a resonance frequency stability of $10^{-13}$ or less [17]. Such sub-μK monitoring cannot be achieved with conventional thermistors due to the thermistor nonlinearity. The temperature of the thermistors is generally approximated by using the Steinhart-Hart thermistor third-order approximation. To overcome the detection and control limit, a cross-polarized dual-mode temperature stabilization method has been suggested in birefringent crystalline resonators [18-20]. The approach draws from the frequency stabilization of quartz oscillators using two mechanical modes with different thermal sensitivities [21]. In our approach and prior studies, the laser frequency is locked to an extraordinary polarized resonator mode, and its radio-frequency modulated sideband, or a second laser frequency, is subsequently locked to an ordinary polarized resonator mode. The different thermorefractive coefficients of the orthogonally polarized modes allow the thermal sensitivity of the resonator mode to be detected down to sub-μK levels with a precisely measurable cross-polarized dual-mode radio-frequency beat, which facilitates the precise thermal stabilization of the resonator.

Previous experiments implementing this approach have reported the microresonators that are temperature-stabilized with 10- to 100-nK precision, which is calibrated by the in-loop cross-polarized dual-mode beat frequency measurement. For the sub-μK resonator temperature, the laser frequency instability locked to such resonators should reach less than $10^{-13}$ even at 1000 s integration time based on the theoretical thermodynamic noise models [22]. However, the frequency instability reported in the experiments shows substantial discrepancy from the theoretical model, implying that the stringent correlation between the resonator temperature and the cross-polarized dual-mode beat frequency could no longer hold true in time. This discrepancy raises the question of whether the thermodynamically limited resonator temperature stabilization using the dual-mode beat stabilization method is fundamentally insufficient for the realistic millimeter-scale ultrahigh-$Q$ resonator or the discrepancy stems from extraneous noise that needs to be identified and compensated.

First, one can consider that this mismatch arises from the temperature nonuniformity or temperature gradient caused by the temperature difference between the resonator and its enclosure or between the substrate and the resonator, leading to the imperfect control of the resonator thermal expansion. Second, the waveguide effects and the modal interactions experienced by optical modes can possibly modify the refractive indices and the mode volumes of the resonator. Third, the cross-polarization metrology probes predominantly the optical mode volume temperature with the optical mode distribution at the resonator rim, while the resonator volume temperature − and hence the volumetric size fluctuation − may be less correlated. In addition, the finite heat diffusion time-scale results in a time lag (before thermal equilibrium) in the resonator stabilization. The first question was studied by Baumgartel *et al.* in Ref. [19]. Varying the temperature difference between the resonator and its enclosure introduces an extraradial deformation caused by the variation in temperature gradient along the radial direction, and the deformation increases with the size of the resonator. In this work, we implement a thin cylindrical ultrahigh-$Q$ magnesium difluoride ($MgF_2$) resonator with a smaller radius by a factor of 2.407 compared to the one used in Ref. [19] to alleviate the impact of the radial deformation caused by the resonator temperature gradient, and investigate the others. Light guiding effects such as refractive index modification, mode extinction, and modal area expansion are further supported by finite-element-method (FEM) numerical simulations on potential effects on the dual-mode temperature stabilization. Moreover, we study the impact of circulating laser-induced heating and heat diffusion with an associated time delay in the resonator.

To interrogate the resonator mode volume temperature, two low-phase-noise continuous wave (cw) lasers are locked to a transverse magnetic (TM) mode and a transverse electric (TE) mode of the whispering-gallery resonator, and the dual-mode beat frequency is measured by heterodyning the two stabilized lasers at a photodetector. We measure and demonstrate the time-dependent frequency correlation of the dual-mode beat frequency with the resonator temperature, represented by the resonant frequency drift due to the thermal expansion, by simultaneously counting the dual-mode beat frequency and a polarization-mode stabilized laser frequency. The degree of correlation between the open-loop dual-mode beat frequency and the resonance frequency approaches the theoretical bounding limit, where the averaged resonator volume temperature change is equal to the optical mode volume temperature change, with increasing integration time. By locking the dual-mode beat frequency to a radio-frequency clock with a passive suppression of the extraneous noise, we achieve a 51.78× enhancement in long-term frequency instability down to 14.55 kHz at 1000 s integration time, corresponding to a temperature instability of 8.53 μK and a linear drift of 0.54 kHz min$^{-1}$.

**Results**

**Theoretical model of dual-mode temperature stabilization**

The resonator is made from a *z*-cut magnesium difluoride ($MgF_2$) single crystal. It is a thin disk-shaped resonator with a radius (*R*) of 1.35 mm, which is sizably smaller than that of our prior study to mitigate the temperature gradient and shorten the heat diffusion time to achieve fast thermal equilibrium. A ring-down time of 3.63 μs is measured, corresponding to *Q* of $2.1\times10^9$, as illustrated in Figure 1a. MgF$_2$ is a birefringent crystal with $n_e = 1.382$ and $n_o = 1.37$, where (*o*) represents the ordinary polarized light or polarized with a TM mode and (*e*) represents the extraordinary polarized light or polarized with a TE mode, for the *z*-cut resonator. The resonance frequency in the whispering-gallery resonator has a linear function of temperature dependence that can be quantified using the thermal expansion coefficient, describing the resonator expansion perpendicular to the crystal axis ($\alpha_{l,(o)}$), and thermorefractive coefficients ($\alpha_{n,(o,e)}$). The expansion along the symmetry axis has a much weaker impact on the mode frequency and can be neglected. At room temperature, these coefficients are $\alpha_{l,(o)} = (1/R)(dR/dT) = 8.9\times10^{-6}$, $\alpha_{n,(e)} = (1/n_e)(dn_e/dT) = 0.64\times10^{-6}$, and $\alpha_{n,(o)} = (1/n_o)(dn_o/dT) = 0.23\times10^{-6}$ [23].

The thermal dependence of the TM and TE mode frequency difference, $f_{TM-TE} = f_o - f_e$, (i.e., heterodyne beat note), can be determined by $\Delta f_{TM-TE}/\Delta T = -v_0(\alpha_{n,(o)} - \alpha_{n,(e)})$, where $v_0$ is the optical carrier frequency. To derive this expression, we assume that the modes occupy the same mode volume and are nearly degenerate so that an increase in resonator size does not change their relative frequency. For an optical carrier wavelength of 1565.5 nm, we characterize the dual-mode beat frequency thermal fluctuation (i.e., $\Delta f_{TM-TE}$) in our MgF$_2$ resonator, which is 78.71$\Delta T_m$ MHz based on the difference between the TM and TE mode thermorefractive coefficients, where $\Delta T_m$ is the temperature change in the whispering-gallery mode channel. On the other hand, the frequency shift induced by thermal expansion is described by $v_0 \alpha_{l,(o)}$ for the ordinary polarized mode (or the electric field of the mode is perpendicular to the crystal axis) in the resonator. The coefficient $\alpha_{l,(o)}$ is typically greater than $\alpha_{n,(o,e)}$. The value of the frequency fluctuation of the ordinary polarized mode is estimated to be 1.705$\Delta T_R$ GHz based on the thermal expansion coefficient, where $\Delta T_R$ is the resonator volume temperature change. We note that the temperature fluctuations are quantified with precision frequency metrology. The ratio of the dual-mode beat frequency variation and a single mode frequency variation due to the thermal expansion becomes a factor of 21.66 when the mode volume temperature variation ($\Delta T_m$) is equal to the resonator volume temperature variation ($\Delta T_R$). This correlation can be utilized for stabilizing the resonator temperature by locking the radio-frequency dual-mode beat to a low-noise radio-frequency clock with proper feedback actuation.

**Integration of waveguide effects on the resonator temperature stabilization**

To detail the impacts of waveguide modal effects on the resonator temperature stabilization, we perform the FEM modeling of the MgF$_2$ resonator with a 12.5 μm radius of curvature and a 1.35 mm bending radius. The simulations clearly show the well-confined TM and TE modes at our experimental laser frequencies. The purities of these two modes are high enough, and their extinction ratios from the orthogonal modes are 35 dB (Supplementary Information S1). The eigenfrequencies of the TE and TM modes are determined to be $f_{TE}$ = 191.44842724346256 THz and $f_{TM}$ = 191.45602707701247 THz, respectively. The separation of the two frequencies ($f_{TM-TE}$) is 7.599 GHz, which is close to the experimentally observed value of 5.953 GHz. The deviation is attributed to slight differences in the waveguide geometry between the numerical model and the fabricated device, as well as in minute material constant variations.

We calculate the effective refractive indices including the waveguide modal effect for both TM and TE resonant modes in the resonator. The FEM simulation shows that the modal effect modifies the refractive indices and the resulting refractive indices decrease by a factor of $5\times10^{-3}$ for both modes, as illustrated in Figure 1c. This is small because the modal areas for both confined modes are fairly large (modal area of 75.52 μm$^2$ for the TM mode and 74.56 μm$^2$ for the TE mode) and the bending radius is also large. The refractive index change at the different temperature ($T+\Delta T$) is computed by using temperature-dependent refractive index equations: $\Delta n_e \cong \Delta T [0.09797-5.57293\times10^{-4} T]\times10^{-5}$ and $\Delta n_o \cong \Delta T [0.04183-5.63233\times10^{-4} T]\times10^{-5}$ from Ref. [17]. The thermorefractive index changes due to the temperature change are small – we use a $\Delta T$ = 10 K to visualize the modification. Regardless of the temperature (even with $\Delta T$ = 10 K), the refractive index modification due to the waveguide modal effect is equally a factor of $5\times10^{-3}$ for each mode, implying that the difference in refractive indices [$\alpha_{n,(o)} - \alpha_{n,(e)}$] can thus be safely considered the same for our dual-mode temperature stabilization. We also simulate the possible anisotropic thermal expansion of the mode cross-sections. For 1 K of the temperature change, the resonator deformation is small enough, as illustrated in Supplementary Information S2, and the distinct mode area change is not observed. If we consider that our resonator mode volume temperature can be stabilized to sub-µK, the modal size fluctuations can therefore be effectively neglected for stabilization.

**Correlation measurements between the dual-mode frequency and the resonator temperature**

We encapsulate the microresonator in a compact aluminum oven with temperature control and photodetector units in a small form factor, as shown in Figure 1b. The resonator temperature is prestabilized at 295.7 K with mK precision over hours via the proportional-integral-differential (PID) control of a thermoelectric cooler (TEC) in a resonator mount. The resonator is then placed in a compact vacuum chamber, and the chamber is evacuated to maintain a high vacuum level ($8\times10^{-6}$ torr) to eliminate convective heat transfer and environmental perturbations such as pressure and humidity. For instance, strong acoustic noise peaks are eliminated from the dual-mode beat phase noise, as illustrated in Supplementary Information S3. Figure 2 shows the schematic experimental measurement set-up. We utilize two cw lasers to interrogate a TM mode and a TE mode, respectively. The lasers that we used for stabilization to the resonator have a linewidth of 3 kHz at 100 ms integration time. The laser helps to interrogate ultrahigh-$Q$ resonances without unwanted heating of the resonator so that the short-term scale temperature stability is not limited by the laser. The orthogonally polarized laser beams are combined with a fiber-optic polarization beam combiner and evanescently coupled into the whispering-gallery resonator via a prism [24].

Each laser frequency is locked to the resonance mode via the Pound-Drever-Hall (PDH) technique [25]. The optical power used to interrogate the resonator is 20 µW for each resonance mode. We first find the resonance frequency of a TM mode and then interrogate a TE mode near the TM mode, offering the best PDH error signals. A single photodetector is implemented, and the two PDH error signals are separated by two different electro-optic modulation frequencies.

The frequency difference in TM and TE modes is deterministically measured at 5.953 GHz. This dual-mode beat frequency is divided by 64 with a prescaler for stabilization to a 93.015 MHz radio-frequency clock. We then examine the Allan deviation (AD) of the TM laser frequency and the dual-mode beat frequency to measure the degree of correlation in time. The stability of the TM laser frequency is measured by downconverting the optical frequency via the heterodyne beating of the TM laser against a commercial fiber laser frequency comb locked to an ultrastable laser possessing 1 Hz linewidth and ≈ 1 Hz s$^{-1}$ drift [26]. Figure 3a shows both the TM laser frequency ($f_{TM}$) measurement data and the dual-mode beat frequency ($f_{TM-TE}$) measurement data with 1 s counter gate time. Figure 3b plots the calculated AD from the frequency measurement datasets. For the open-loop measurement, both ADs show frequency drift over time. The long-term stability of 750 kHz at 1000 s integration time implies less than mK resonator temperature stability dominated by the thermal expansion coefficient. We extract the correlation between $\Delta f_{TM}$ and $\Delta f_{TM-TE}$ along the integration time (τ) by measuring their ratio as expressed by: $\beta(\tau) = \Delta f_{TM}(\tau)/\Delta f_{TM-TE}(\tau)$.

Figure 3c shows that $\beta(\tau)$ approaches the theoretically estimated value of 21.66 when $\Delta T_R = \Delta T_m$ with increasing integration time, which indicates that the resonator volume temperature equilibrates to the mode volume temperature via heat diffusion and time averaging. This is supported by a heat transfer simulation with circulating laser power as a heat source (details in Supplementary Information S4). The simulation shows a heat gradient along the radius due to the light absorption by the resonator in Figure 3d. The average temperatures near the mode volume and the entire resonator volume are probed individually and their difference is calculated in time, as illustrated in Figure 3e. The difference between the mode volume temperature and the resonator volume temperature becomes constant after 30 s, which could be the characteristic time required to reach heat equilibrium. The longer time averaging helps $\beta(\tau)$ to be closer to the theoretical value. The deviation of $\beta(\tau)$ from the theoretically estimated value of 21.66 is attributed to the atmospheric temperature variation triggering the TEC control to maintain the enclosure temperature, which can be an extra heat source for the resonator.

**Dual-mode temperature stabilization and residual drifts**

We stabilize the dual-mode beat frequency by controlling the intensity of the TE laser into the resonator with an acousto-optic modulator (AOM). The inset in Figure 4a shows the frequency measurement of the in-loop dual-mode beat carrier locked to a radio-frequency reference with 1 s gate time. While the dual-mode beat frequency is locked, we count the TM laser frequency and plot the measured AD, as shown in Figure 4a (navy scatters). At the characteristic heat equilibrium time, the largest stability enhancement is accomplished, as shown in Figure 4b. In this regime, further noise suppression is limited by the feedback gain for the dual-mode beat control to avoid the electrical-line harmonic noise peaks on top of the uncompensated frequency noise. For $\tau < 30$ s, the stability enhancement could be limited by the response of the resonator temperature to the feedback control so that the magnitude of the enhancement decreases. For $\tau > 30$ s, we observe a linear frequency drift in AD, leading to the roll-off of the stability enhancement in Figure 4b.

We investigate this frequency drift by measuring and analyzing the frequency noise of the dual-mode beat frequency and find that the drift could be attributed to the noise of the radio-frequency clock to which the dual-mode beat frequency is locked. Figure 5a illustrates the stabilized dual-mode beat frequency spectrum (inset) and its single-side-band (SSB) phase noise. The dual-mode beat frequency stabilization loop suppresses noise up to 300 Hz offset frequency. When the higher feedback gain was applied, we found strong 60 Hz harmonic peaks in the error signal, which limits the available feedback gain. However, we found that the feedback bandwidth and gain could not be the fundamental limiting factors for the long-term temperature stability of the resonator. In principle, time averaging with more than the characteristic time, which is required to reach the resonator heat equilibrium ($\approx 30$ s), can help to rule out fast frequency fluctuations. The dual-mode beat frequency noise starts to converge to the reference clock noise below 1 Hz offset frequency. The locked dual-mode beat frequency SSB phase noise has a decaying slope of $f^{-1}$ while the radio-frequency reference SSB phase noise has a decaying slope of $f^{-4.6}$ near the carrier frequency, which implies that the two curves could have a crossover so that the achievable dual-mode beat frequency stability can be bounded by the radio-frequency clock noise beyond the crossover point, which potentially introduces the frequency drift in our current system. To verify this assumption, we deliberately reduce proportional feedback gain without losing frequency locking and measure the stability again. Although the stabilities at $\tau <$ 100 s are rather worse, the frequency drift is significantly suppressed, as illustrated in Figure 5b. The drifts estimated with a linear function fitting show 11.8 kHz min$^{-1}$ and 0.54 kHz min$^{-1}$ at the two different gains. The AD at 1000 s integration time is improved by an order-of-magnitude compared to that with the higher gain and the resonator temperature remains to be approximately 10 μK from 1 to 1000 s integration time. The TM laser frequency instability is improved approximately by a factor of 51.78 to 14.55 kHz at 1000 s integration time compared to the instability (753.8 kHz) before stabilizing the dual-mode mode beat frequency as illustrated in Figure 5d.

**Residual intensity noise contributions**

We investigate the impact of the coupled laser intensity fluctuation resulting in the fluctuation of the absorbed circulating optical power via the amplitude fluctuation of the intracavity laser field. Since we implement the TE laser intensity modulation feedback, the intensity fluctuation of the TE laser into the resonator is under control, but the intrinsic relative intensity noise (RIN) or the coupled power change of the TM laser could trigger the intracavity laser field fluctuation. To quantify the impact of RIN, we measure the resonance frequency shift induced by the laser intensity modulation as illustrated in Supplementary Information S5. The coupled laser power is modulated by an AOM with a 1 Hz top-hat function and the modulated laser power (0.75 %) into the resonator and the corresponding TM laser frequency shift are measured. We determine the resonance frequency shift on the coupling power modulation – 2.66 kHz for 1 % coupled power change corresponding to 200 nW. The integrated RIN of the lasers used in this measurement is ≈ 10 ppm from 10 Hz to 10 MHz offset frequencies [27]. By extrapolating RIN to the carrier frequency with 20 dB/decade, we estimate that the impact of RIN could be one part in $10^3$ up to 1 Hz offset frequency. Hence, the RIN-associated power fluctuation can introduce a TM laser frequency fluctuation of 266 Hz, which is small compared to our currently measured frequency instability, but the laser RIN or the coupling power fluctuation could be a contributing bound when operating at the thermodynamic noise limit.

**Discussion**

We implement the cross-polarized dual-mode temperature stabilization for a birefringent high-$Q$ whispering-gallery resonator and improve the long-term stability by ≈ 51× at 1000 s integration time. We achieve 10 μK resonator temperature instabilities even up to 1000 s integration time, enabling this compact optical resonator module to serve as a high-performance frequency reference in potential metrology, timing and frequency transfer applications. The numerical simulations reveal the possible thermorefractive coefficient and mode volume modification due to the waveguide effect and the heat diffusion process in the $MgF_2$ resonator. The characteristic time for the resonator temperature equilibrium is deduced and matched with the experimental result. Further, more accurate models consider the impact of TEC control because the ambient temperature variations can trigger the TEC to maintain the enclosure temperature, leading to a change in the resonator temperature. Improved system thermal isolation from the surrounding environment may be needed. Although the current long-term frequency or temperature stability seems to be restricted by the low Fourier frequency noise of the radio-frequency clock used for the dual-mode beat frequency stabilization, further investigations and understandings on noise are desirable for operating the resonator stability at fundamental thermodynamic noise limits.

**Materials and Methods**

**FEM optical and thermal modeling and simulations for the MgF$_2$ resonator**

The thin cylindrical MgF$_2$ resonator is modeled using 2D-axis symmetry in COMSOL multiphysics with a 1.35 mm bending radius and a 12.5 μm radius of curvature. The guided modes in the resonator are calculated by solving Maxwell's electromagnetic wave equations in the frequency domain, which provides the eigenfrequencies of the modes. By performing mode analysis near the optical carrier frequency, the family of TE and TM modes associated with the eigenfrequencies are found. We probe the effective modal areas and the effective mode indices, and study their resonator temperature dependence by performing simulations at different resonator temperatures. For the resonator heat diffusion modeling and simulation, the time-dependent heat transfer equation solver in COMSOL is implemented, where a user-defined heat source, the resonator absorption of circulating laser power in the modal area, is used. The size of the heat source is estimated by the modal area calculated by the mode analysis study. The physical constants and the optical parameters used in the simulation are listed in Supplementary Information S4.

**Acknowledgments**

We appreciate the helpful discussions with Shu-Wei Huang, Wenting Wang, Yoo Seung Lee, and Abhinav K. Vinod. The authors acknowledge support from DARPA and Air Force Research Laboratory under contract FA9453-14-M-0090.

**Conflict of interest**

The authors declare that they have no conflict of interest.

**Contributions**

J. L., C.W.W., A.B.M. and L.M. designed the experiment, and J.L. developed the experimental set-up, performed the experiment and analyzed the measurements. W.L. and A.A.S developed the optical resonator along with the package assembly. J.L. performed the FEM numerical simulations and analyzed them with A.B.M. and C.W.W. All authors contributed the manuscript preparation.

Microresonators, and Beam Control XVI. San Francisco, California, United States: SPIE, 2014; 89600X, doi: 10.1117/12.2044824.

**Figure legends**

**Figure 1.** Integrated thin-disk MgF$_2$ ultrahigh-Q whispering-gallery resonator and FEM simulations of the MgF$_2$ whispering-gallery resonator modes.

**Figure 2.** Orthogonally polarized dual-mode temperature stabilization set-up.

**Figure 3.** Correlation between a resonator-stabilized TM laser frequency ($f_{TM}$) and the dual-mode beat frequency ($f_{TM-TE}$).

**Figure 4.** Resonator temperature stabilization and stability enhancement.

**Figure 5.** Single-sideband (SSB) phase noise of the dual-mode beat frequency, Allan deviation, and their enhancement at different proportional feedback gains.

**Figures**

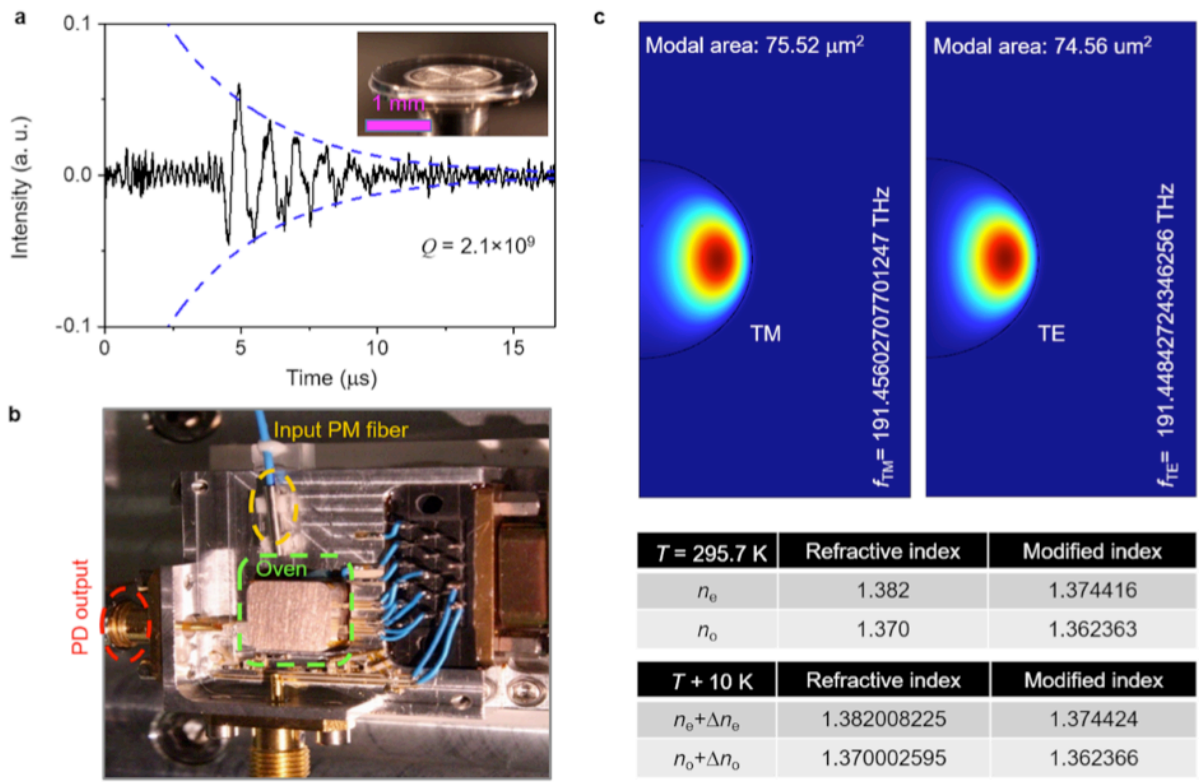

**Figure 1. Integrated thin-disk MgF$_2$ ultrahigh-$Q$ whispering-gallery resonator and FEM simulations of the MgF$_2$ whispering-gallery resonator modes. a** Ring-down measurement of the resonator shows a cavity lifetime of 3.63 μs corresponding to $Q = 2.1 \times 10^9$. Inset: image of the thin-disk whispering-gallery resonator. **b** Integrated package. The resonator is placed in a compact aluminum oven with a PID controlled TEC and the laser light is delivered into the resonator via an optical fiber. Both TM and TE laser signals are measured at a single photodetector. Optical resonance mode frequencies for TM and TE are located at $f_{TM} = 191.45602707701247$ THz and $f_{TE} = 191.44842724346256$ THz. **c** The frequency difference between the two orthogonal modes ($f_{TM-TE}$) is evaluated to be 7.599 GHz, close to the experimentally observed value. The FEM simulation shows modal areas and effective index changes due to the whispering-gallery resonator guiding effect for individual modes. The waveguide effect lowers the refractive indices by a factor of $5 \times 10^{-3}$ for both polarization modes, numerically examined across a $\Delta T$ of 10 K

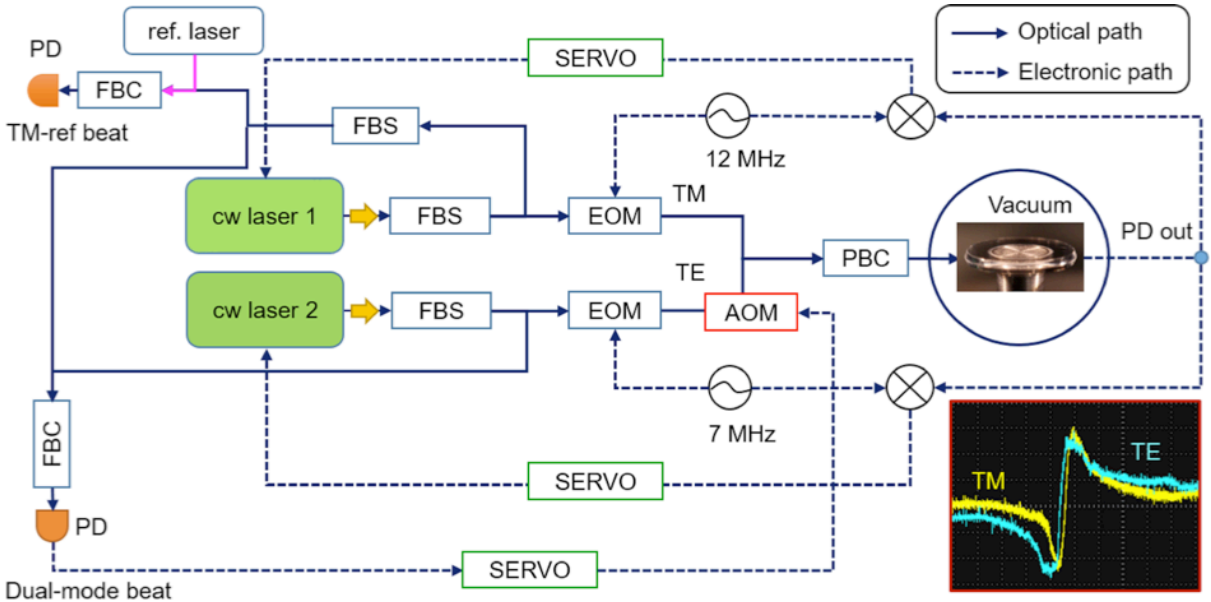

**Figure 2. Orthogonally polarized dual-mode temperature stabilization set-up.** The two cw lasers are coupled into the ultrahigh-Q MgF$_2$ resonator in a vacuum chamber with two orthogonally polarized modes. Each laser is locked to the resonance via a Pound–Drever–Hall (PDH) lock scheme. The inset (right bottom) shows the PDH error signals for the TM and TE modes interrogated by the two lasers, respectively. The dual-mode beat frequency is measured at a single photodetector and the TM laser frequency stability is measured by the heterodyne

beating of it against an ultrastable Fabry–Pérot mirror cavity laser reference possessing 1 Hz linewidth and ≈1 Hz s$^{-1}$ drift. The dual-mode beat frequency is locked to a radio-frequency clock via intensity modulation with an acousto-optic modulator (AOM). Both the TM laser frequency and the dual-mode beat frequency are simultaneously counted and their phase noises are recorded. FBC fiber-optic beam combiner, FBS fiber-optic beam splitter, EOM electro-optic modulator, PBC polarization beam combiner, PD photodetector

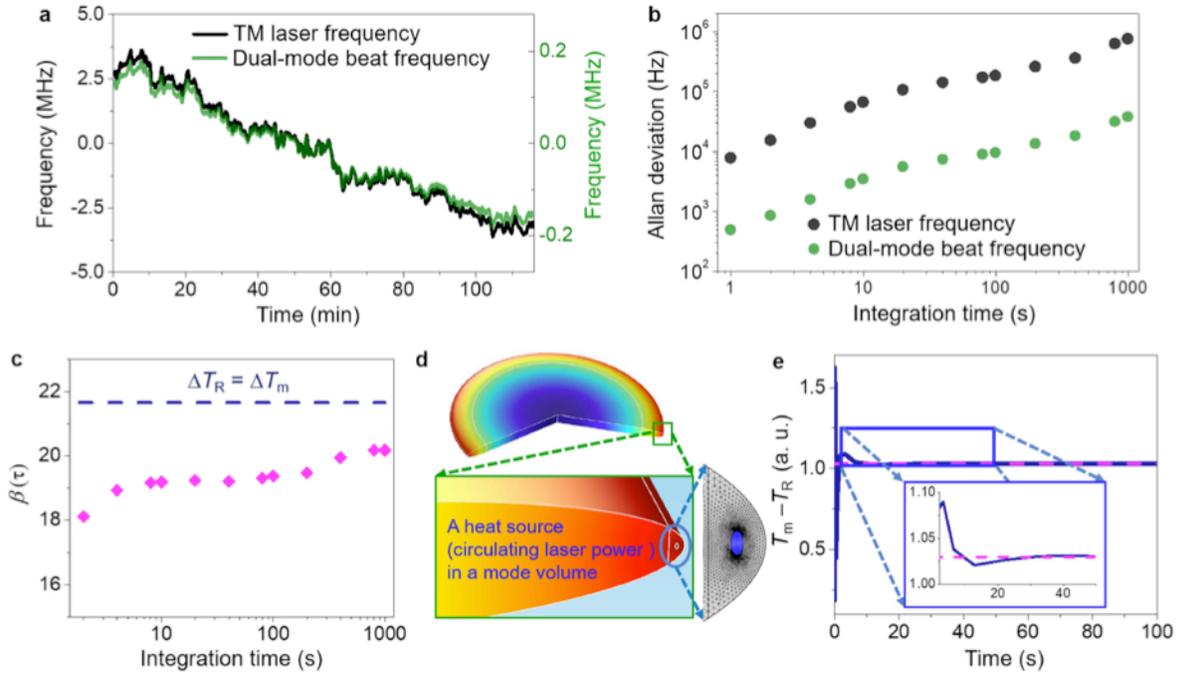

**Figure 3. Correlation between a resonator-stabilized TM laser frequency ($f_{TM}$) and the dual-mode beat frequency ($f_{TM\text{-}TE}$). a** The stabilized TM laser frequency and the dual-mode beat frequency are simultaneously counted for 2 h, with a correlation in the frequency changes. **b** The calculated Allan deviations without the dual-mode temperature stabilization show the frequency drift in time. The long-term stability of 0.75 MHz at 1000 s integration time shows sub-mK resonator temperature stability dominated by the thermal expansion coefficient. **c** The measured $\beta(\tau)$, i.e., $\Delta f_{TM}/\Delta f_{TM-TE}$, shows that the frequency correlation approaches the theoretically estimated value of 21.66 at $\Delta T_R = \Delta T_m$ with increasing integration time, likely because the resonator volume temperature equilibrates to the mode volume temperature via heat diffusion and averaging. **d** The simulated resonator volume temperature distribution at 100 s due to a circulating laser heat source in the optical mode volume. **e** The simulation result shows the time-dependent difference between the mode volume temperature ($T_m$) and the averaged resonator volume temperature ($T_R$). The difference becomes constant after 30 s, which can be the characteristic heat diffusion time required to reach an equilibrium point

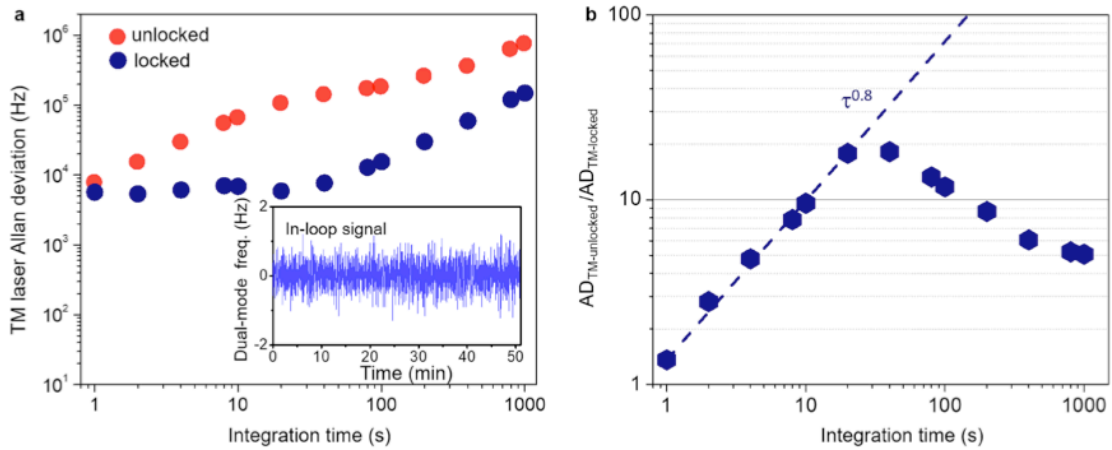

**Figure 4. Resonator temperature stabilization and stability enhancement. a** TM laser frequency stability when the dual-mode beat frequency is unlocked (red) and locked (navy). Inset: in-loop dual-mode beat frequency measurement showing less than ±1 Hz deviation. **b** At the characteristic equilibrium time, the largest stability enhancement is achieved. For $\tau < 30$ s, the stability enhancement is likely to be limited by the response of the resonator temperature to the feedback control. For $\tau > 30$ s, we measure the roll-off of the stability enhancement due to the frequency drift

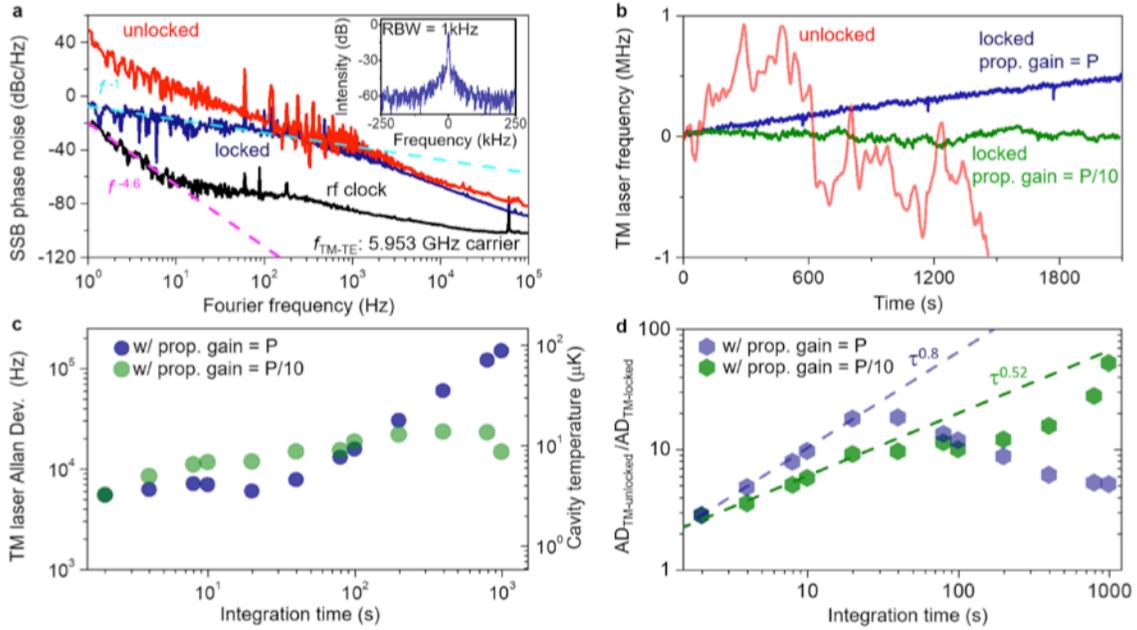

**Figure 5. Single-sideband (SSB) phase noise of the dual-mode beat frequency, Allan deviation, and their enhancement at different proportional feedback gains. a** SSB phase noise measurements of the dual-mode beat frequency at 5.953 GHz. Inset: the stabilized dual-mode beat frequency spectrum at 5.953 GHz. **b** Measured TM laser frequency in time with the two different feedback gains. The frequency drifts estimated with a linear function fitting show two examples at 11.8 kHz min$^{-1}$ (navy) and 0.54 kHz min$^{-1}$ (green), respectively. The red line is the unlocked (open-loop) TM laser frequency for comparison. **c** TM laser frequency Allan deviations with different feedback gains. By lowering the proportional gain, the long-term stability is improved, illustrating that the radio-frequency reference noise could be responsible for the frequency drift. **d** The lower proportional gain makes the enhancement slower along the integration time, but the enhancement roll-off is avoided with the higher gain. The enhancement factor at 1000 s integration time is 51.78

# Supplementary Information

**Probing 10 μK stability and residual drifts in the cross-polarized dual-mode stabilization of single-crystal ultrahigh-*Q* optical resonators**

Jinkang Lim[1,*], Wei Liang[2], Anatoliy A. Savchenkov[2], Andrey B. Matsko[2,*], Lute Maleki[2], and Chee Wei Wong[1,*]

* Corresponding authors: jklim001@ucla.edu; andrey.matsko@oewaves.com; cheewei.wong@ucla.edu

[1] Fang Lu Mesoscopic Optics and Quantum Electronics Laboratory, University of California, Los Angeles, CA 90095 USA

[2] OEwaves Inc., 465 North Halstead Street, Suite 140, Pasadena, CA 91107 USA

## S1. Mode extinction at each polarization mode

There are no pure TM and TE modes but there exist TM-like and TE-like modes. In order to check the purities of the modes, we simulated the both TM and TE modes at each eigenfrequency of the whispering-gallery resonator and the power extinction ratios between TM and TE modes are ≈35 dB, which is pure enough, so that we ignore modal interaction contributions from the other residual mode at each eigenfrequency.

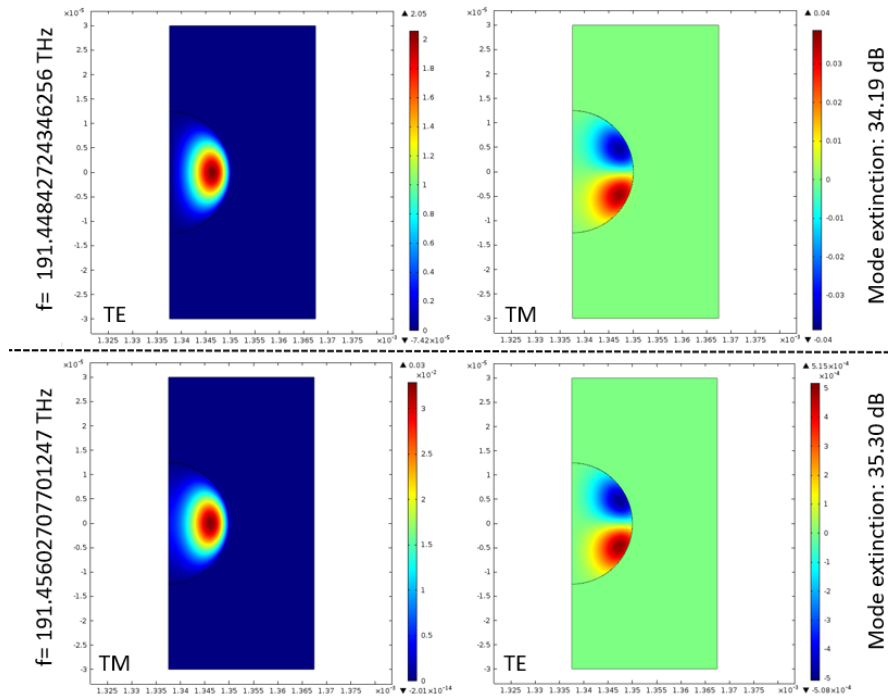

**Figure S1. TM and TE modes extinctions.** The mode extinction for both eigenfrequencies are ≈35 dB.

## S2. Impact of the modal area change due to the thermal expansion

The change in the waveguide dimension due to the thermal expansion could introduce some changes in the size of modal areas ($A_m$) and mode volumes ($V_m$). To estimate this effect, we performed the FEM simulation and derived the change in $A_m$. The different thermal expansion coefficients can change the waveguide shape asymmetrically (i.e. elliptic). The 1 K temperature variation changes the length perpendicular to the optical axis by 222.5 pm and the length parallel to the optical axis by 342.5 pm in the modal area. The deformation is small enough if we consider that our stabilized resonator mode volume temperature is controlled at 10 µK and thus it will not make meaningful impact on the size of the modal area and therefore on the power spectral density of the thermorefractive frequency noise.

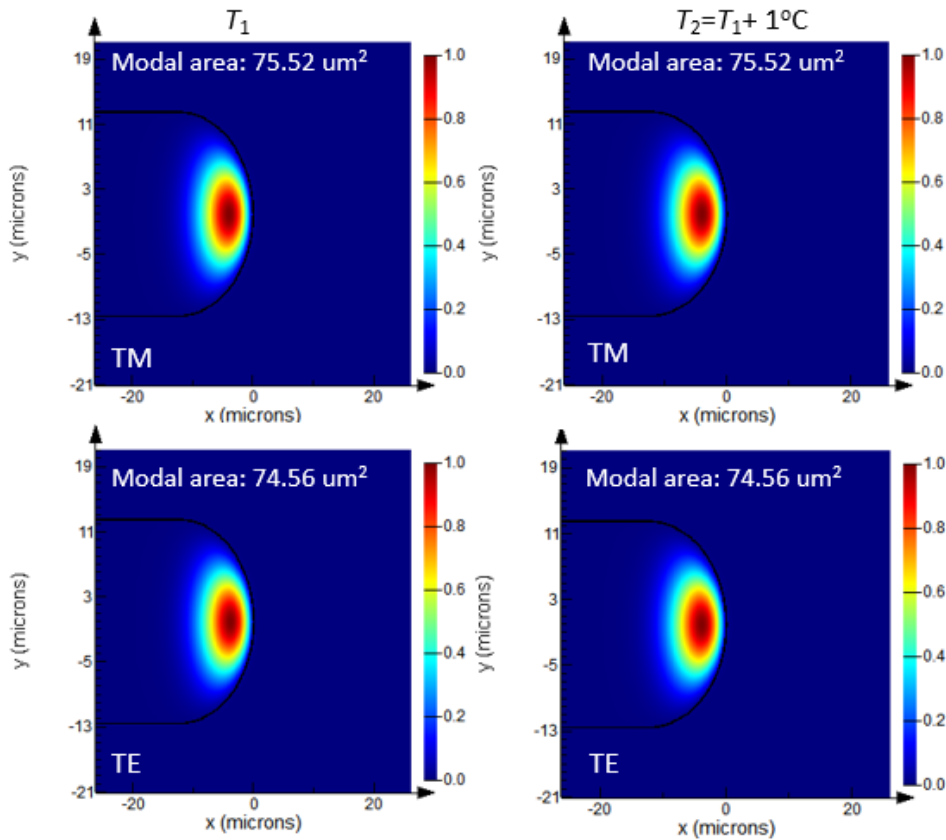

**Figure S2. FEM simulation of modal area variation due to the thermal expansion.** The different thermal expansion coefficient can change the waveguide shape asymmetrically (ellipse). The 1 K temperature variation changes the resonator length perpendicular to the optical axis by 222.5 pm and the resonator length parallel to the optical axis by 342.5 pm in the modal area. The deformation is small enough and does not make meaningful impact on the size of the modal area in numerical simulations.

## S3. Impact of ambient pressure

The microresonator is placed in a compact vacuum chamber and evacuated to eliminate convective heat transfer and ambient perturbations such as pressure and humidity. The strong acoustic noise peaks above 1-kHz offset frequency are attenuated or eliminated, which is not compensated by the active dual-mode temperature stabilization feedback loop.

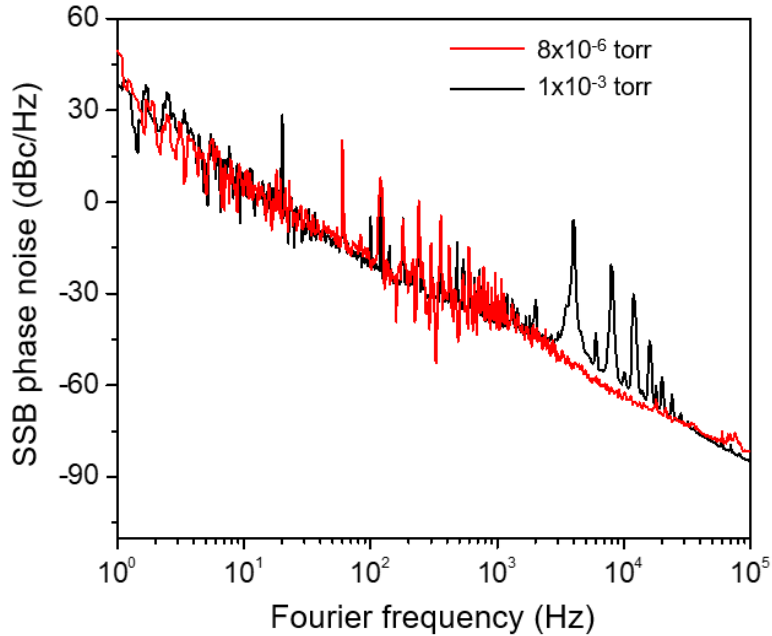

**Figure S3. SSB phase noise of the dual-mode beat frequency in different pressure levels.** Red line at $8\times10^{-6}$ torr and black line at $1\times10^{-3}$ torr. Acoustic noise is further suppressed at the lower pressure level.

## S4. Heat diffusion in the resonator and the resonator volume temperature distribution

The absorbed light by the resonator can change not only local mode volume temperature but also change the resonator volume temperature via heat diffusion. We simulate the heat diffusion in the resonator caused by a laser heat source with the time-dependent heat diffusion analysis tool in the COMSOL Multiphysics, in which we exclude the contribution of independently feedback-controlled TEC. The elliptic-shaped heat source is centered at the TE mode peak with $\approx 10$ μm$^2$ because most laser energy is concentrated near the mode peak as illustrated in Figure S4a and the 2D axis-symmetry geometry are implemented for simulating the cylindrical whispering-gallery resonator structure. We use parameters that use in the experiment for this simulation and they are shown in a table. The thermal conductivity of 21 W•m$^{-1}$K$^{-1}$, the density of 3180 kg• m$^{-3}$ and heat capacity at constant pressure of 920 J•kg$^{-1}$K$^{-1}$ are used for the MgF$_2$ resonator respectively. We calculate a characteristic diffusion time (≈30 sec) that provides the difference of the mode volume temperature ($T_m$) and the averaged resonator volume temperature ($T_c$) becomes a constant illustrated in Figure S4b and the result is also shown in Figure 3e. In the simulation, $T_m$ is measured around the optical mode volume (circled area in Figure S4a) and $T_c$ is measured for the entire resonator.

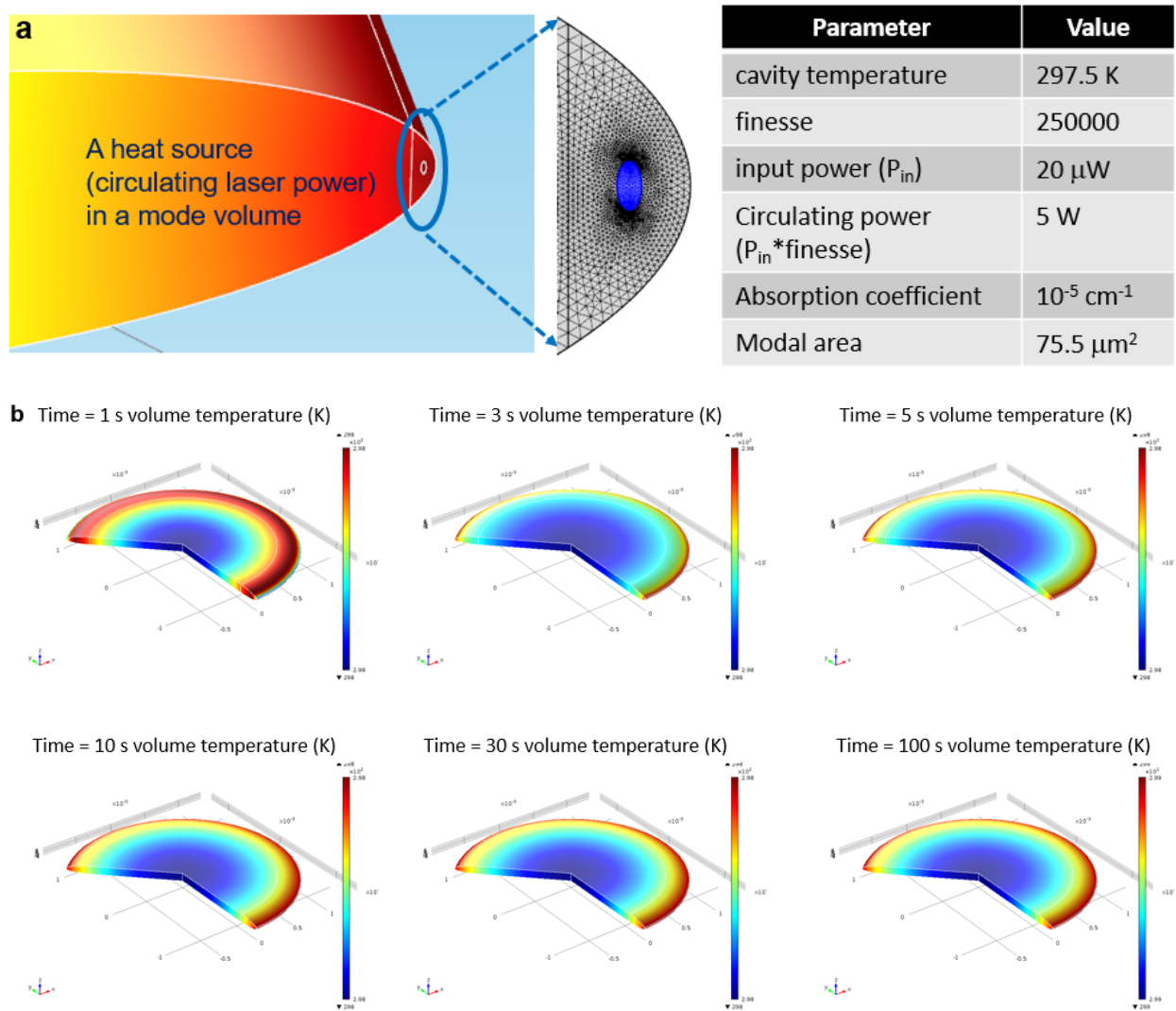

**Figure S4. FEM simulation of the heat diffusion in the resonator.** FEM simulation of the heat diffusion in the resonator and the resonator volume temperature distribution in time when a laser heat source is located at the center of the TE mode. The temperature distribution shows negligible changes after 30 s.

**S5. Impact of the relative intensity noise (RIN) on the resonator temperature stability**

To quantify the laser RIN contribution to the resonator temperature stability. We investigate a transfer function showing the resonance frequency shift induced by the laser intensity modulation. The coupled laser power is modulated by an acousto-optic modulator with a 1 Hz top-hat function with the modulated laser power (0.75 %) into the resonator and the corresponding TM laser frequency is measured. TM laser frequency shift is measured. We determine resonance frequency shift on the coupling power modulation – 2.66 kHz for 1 % coupled power change corresponding to 200 nW.

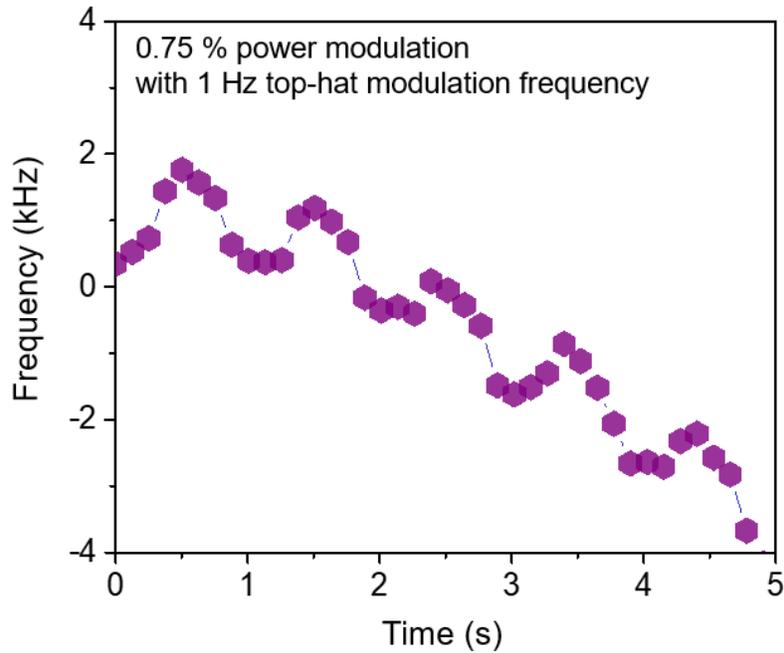

**Figure S5. The TM laser frequency shift due to the coupled laser intensity modulation.** The TM laser frequency response to the input power modulation. The TE laser is modulated with a 1 Hz top-hat function and the TM laser frequency response is measured to be 2 kHz for 0.75 % coupled power change.